\documentclass[showpacs,preprintnumbers,amsmath,amssymb]{revtex4}
\usepackage{graphicx}
\usepackage{verbatim}
\usepackage{graphicx}
\usepackage{verbatim,hyperref}
\usepackage[cp1251]{inputenc}
\usepackage[english, russian]{babel}
\usepackage{epsfig,psfig}

\begin{document}

\title{Cosmological evolutions of completely degenerated
Fermi-system with the scalar interparticles interactions}

\author{Yu.G.Ignatyev, R.F. Miftakhov}
\email{rustor@bk.ru} \affiliation{Department of Mathematics, Kazan
State University, Mezhlauk 1 str., Kazan 420021, Russia}

\begin{abstract}
Cosmological solutions of Einstein's equations for equilibrium
statistical systems of particles with scalar interaction are
investigated. It is shown that the scalar field can effectively
change the state equation of a statistical system, that leads to
the possibility of secondary acceleration of the cosmological
expansion.
\\
\end{abstract}


\maketitle

\section{Introduction}
To the moment it has been published many papers devoted to the
secondary acceleration of the Universe. In order to solve the
problem of the secondary acceleration of the Universe it is often
proposed in many papers to sufficiently change the fundamental
principles of physics. However, recently there are indications
that complicated multi-component, classical physically systems can
also bring to the secondary acceleration of the
Universe\footnote{In particular, such arguments were represented
by Dmitry V. Gal’tsov, as well as present paper's Authors at
Gracos-2009}. In this case there is no need to revise the
fundamental principles of physics. Some indications of the
possibility of such behavior of the systems with scalar ineraction
were given in the paper \cite{YuMif}. Scalar fields were
introduced in general relativistic statistics by one of the
Authors in the early 1980 years \cite{Yukin3}, \cite{Yukin1}. In
these papers, based on the kinetic theory, a system of particles
with scalar interaction is obtained.

\section{Macroscopic densities}
The full degeneration condition supposes:
\begin{equation}\label{1}
\frac{\mu}{\theta}\to\infty.
\end{equation}
($\mu$ - chemical potentials, $\theta$ - temperature) local
equilibrium distribution function of Fermi-system  is a step
function (see for example \cite{Land}).  In this case, the local
equilibrium distribution function has the form \cite{Yukin1}:
\begin{equation}\label{2}
f^0(x,P)=\chi_+(\mu-\sqrt{m_*^2+p^2}),
\end{equation}
where $\chi_+(z)$ is a Heaviside function;
\begin{equation}\label{2a}
m_*=|m+q\Phi|,
\end{equation}
are the particles effective masses, $q$ is a scalar charge of the
fermions, $\Phi$ is a scalar field potential, $m$ is a vacuum mass
of the fermions. Therefore the integration of macroscopic
densities are representable in elementary functions \cite{Yukin1}:
\begin{eqnarray}\varepsilon_f = \frac{m_*^4}{8\pi^2}
\bigl[\psi\sqrt{1+\psi^2}(1+2\psi^2)~\hskip 60pt \nonumber\\-\ln
(\psi+\sqrt{1+\psi^2})
\bigr]; \label{3a}\\
\label{3b} p_f =\frac{m_*^4}{24\pi^2}
\bigl[\psi\sqrt{1+\psi^2}(2\psi^2-3)~\hskip 60pt \nonumber\\ +3\ln
(\psi+\sqrt{1+\psi^2})\bigr];\\
\label{3c} T_f=\varepsilon_f-3p_f=~\hskip 120pt \nonumber\\
\frac{m_*^4}{2\pi^2}\left[\psi\sqrt{1+\psi^2}-\ln
(\psi+\sqrt{1+\psi^2})\right];
\end{eqnarray}
\begin{equation}\label{3d} \varepsilon_f+p_f=\frac{m_*^4}{3\pi^2}\psi^3\sqrt{1+\psi^2};\end{equation}
\begin{eqnarray}\label{3e} \sigma=\frac{q}{m_*}T_f=~\hskip 60pt \nonumber\\
\frac{q\cdot m_*^3}{2\pi^2}\left[\psi\sqrt{1+\psi^2}-\ln
(\psi+\sqrt{1+\psi^2})\right];
\end{eqnarray}
where
\begin{equation}\label{psi}\psi=p_F/|m_*|
\end{equation}
is a ratio of the Fermi momentum to effective mass,
$\varepsilon_f, p_f, \sigma$ is an energy density, pressure and
scalar charge density of fermions, respectively\footnote{The
specifications see in \cite{Yukin1}.}. In this case the
self-consistent equation of massive scalar field with a mass of
fermions $\mu$ \cite{Yukin1}:
\begin{equation}\label{field1}
\box \Phi + \mu^2 \Phi= -4\pi\sigma.
\end{equation}
The fermion number density is connected with the Fermi momentum by
relation \cite{Land}:
\begin{equation}\label{4}
n(x)=\frac{p^3_f}{3\pi^2}\Rightarrow p_f=(3\pi^2
n(x))^{\frac{1}{3}}.
\end{equation}
Thus, the variable $\psi$ can be expressed in terms of two scalars
- the particles number density, $n(x)$, in the own frame of
reference and the scalar potential, $\Phi(x)$:
\begin{equation} \label{5}
\psi=\frac{(3\pi^2n(x))^{\frac{1}{3}}}{|m+q\Phi|}.
\end{equation}
\section{Self-consistent system of equations for a flat Friedmann model}
Let's consider a cosmological situation where matter is
represented by only a degenerate Fermi system with scalar
interaction of particles and the associated scalar field,
described by the equation (\ref{field1}) (see \cite{YuMif}). As
the metric we choose the Spatially Flat Friedmann Model
\cite{Land1}):
\begin{equation}\label{6}
ds^2=dt^2-a^2(t)(dx^2+dy^2+dz^2),
\end{equation}
in which the independent Einstein equations have the form:
\begin{equation}\label{7}
3\frac{\dot{a}^2}{a^2}=8\pi\varepsilon;
\end{equation}
\begin{equation}\label{8}
3\frac{\dot{a}}{a}=-\frac{\dot{\varepsilon}}{\varepsilon+P}.
\end{equation}
In this metric the conservation law for particles takes form:
\footnote{Everywhere $G=\hbar=c=1$, time and mass are measured in
Planck units.}:
\begin{displaymath}\partial_t \sqrt{-g}n =0,\end{displaymath}
from where, taking into account epression (\ref{4}) we obtain a
momentum integral instead of the energy integral  \cite{YuMif}:
\begin{equation}\label{9}
a p_F=\mbox{Const}.
\end{equation}
Supposing further $\Phi=\Phi(t)$, we obtain
$\varepsilon=\varepsilon(t)$ $p=p(t)$ and  the structure of the
total scalar field EMT in the form of the EMT of a perfect fluid
with the macroscopic velocity $v^i=\delta^i_4$, energy density
$\varepsilon_s(t)$ and pressure $p_s(t)$:
\begin{equation}\label{10}
\varepsilon_s=\frac{1}{8\pi}(\dot{\Phi}^2+\mu^2_s\Phi^2);\quad
p_s=\frac{1}{8\pi}(\frac{1}{3}\dot{\Phi}^2-\mu^2_s\Phi^2).
\end{equation}
Differentiating (\ref{10}) and substituting the results in
(\ref{8}), it can be seen, that this equation is a consequence of
field equations (\ref{field1}), which in metric (\ref{6}) takes
form:
\begin{equation}\label{20}
\frac{1}{a^3}\frac{d}{dt}(a^3\frac{d\Phi}{dt}) +\mu^2
\Phi=-4\pi\sigma.
\end{equation}
Thus, for an independent equations  one can chose  the field
equation (\ref{20}) and one of Einstein's equations (\ref{7}),
which with the condition ($\dot{a}>0$) can be written as:
\begin{equation}\label{field2}
\frac{\dot{a}}{a}=\sqrt{\frac{8\pi}{3}\varepsilon},
\end{equation}
where
\begin{displaymath}
\varepsilon=\varepsilon_f+\varepsilon_s
\end{displaymath}
is a total energy density of fermions and a scalar field.

Let us carry out an analysis of possible cosmological evolution of
the degenerate scalar charged fermions system. From the
expressions for the macroscopic densities (\ref{3a}), (\ref{3b})
we obtain:
\begin{eqnarray}\label{11}
p_f=\frac{1}{3}\varepsilon_f-\hskip 130pt \nonumber\\
\frac{m_*^4}{24\pi^2}\left[4\psi\sqrt{1+\psi^2}-3\ln
(\psi+\sqrt{1+\psi^2})\right].
\end{eqnarray}
We can show, that the expression in square brackets on the right
side (\ref{11}) is a nonnegative for nonnegative $\psi\geq 0$.
Therefore the strict inequality is satisfied:
\begin{equation}\label{11a}
0\leq p_f \leq \frac{1}{3}\varepsilon_f.
\end{equation}
Similarly from (\ref{10}), we find the relation for the scalar
field:
\begin{equation}\label{12}
p_s=\frac{1}{3}\varepsilon_s-\frac{1}{6\pi}\mu^2\Phi^2,
\end{equation}
Also for the scalar field the strict inequality is satisfied:
\begin{equation}\label{12a}
p_s\leq \frac{1}{3}\varepsilon_s.
\end{equation}
Now we find {\it coefficient of barotropic}, as a coefficient
$\kappa$ in the linear relation between total pressure and energy
density of matter:
\begin{equation}\label{kappa}
p=\kappa \varepsilon.
\end{equation}
As a result we find:
\begin{equation}\label{kappa1}\kappa(t)=\frac{1}{3}-
\frac{m_*^4G(\psi)+4\pi\mu^2\Phi^2}{24\pi^2(\varepsilon_s+\varepsilon_f)},
\end{equation}
where:
\begin{displaymath}
G(\psi)=4\psi\sqrt{1+\psi^2}-3\ln (\psi+\sqrt{1+\psi^2}).
\end{displaymath}
So, we can show, that for the concerned system the following
relation is always fulfilled:
\begin{equation}\label{13}\Rightarrow -1 \leq\kappa(t)\leq{\frac{1}{3}}.\end{equation}
Thus, for the cosmological acceleration Universe, $\Omega$,  that
consists of degenerate fermions and scalar field,
\begin{equation}\label{14}
\Omega=\frac{a\ddot{a}}{\dot{a}^2}=-\frac{1}{2}(1+3\kappa),
\end{equation}
we obtain the restriction:
\begin{equation}\label{14a}
-1\leq \Omega\leq 1.
\end{equation}

\section{Numerical solution of Einstein's equations for a degenerate plasma with scalar field}
Attempts of direct numerical integration of Einstein and
Klein-Gordon equations (\ref{7}), (\ref{field1}) in most cases
don't give results owing to nonlinear character of these equations
and ambiguity of radical functions and logarithms in the right
side of equations. Therefore, for the numerical solution of
Einstein's equations the expressions for the macroscopic
densi\-ties were extrapolated by means of elementary functions. As
it can be seen from (\ref{3a}) - (\ref{3e}), all these
expressions, taking into account the integral (\ref{9}) up to
multi\-pli\-ca\-tion by a conformal factor $1/a^4$, --
\begin{equation}\label{15}
m_*^4=\left(\frac{a_0 p^0_F}{a\psi}\right)^4.
\end{equation}
are elementary functions of a single dimensionless function
$\psi$. This fact allows us to find the inter\-po\-la\-tion
expression for the corresponding conformal densities:
\begin{equation}\label{16}
\tilde{p}_f=a^4 P_f;\quad
\tilde\varepsilon_f=a^4\varepsilon_f;\quad
\tilde{\sigma}=a^4\sigma.
\end{equation}
Interpolation of analytic functions for the energy density,
pressure and charge density on the range under consideration can
be written as:
\begin{equation}\label{17}
\tilde{p}_f=\frac{8}{15}e^{-2\psi}\psi+\frac{2\psi^2}{3(1+\psi^2)}
\end{equation}
\begin{equation}\label{18}
\tilde\varepsilon_f=\frac{\left(1-e^{-2\psi-2\psi^2}\right)\left(\frac{38}{15}+
\frac{4}{3\psi^2}+\frac{8\psi^2}{15}\right)}{1+\frac{4\psi^2}{15}}
\end{equation}
\begin{eqnarray}\label{19}
\tilde{T}_f=\frac{\left(1-e^{-2 \psi-2 \psi^2}\right)
\left(\frac{38}{15}+\frac{4}{3 \psi^2}+\frac{8
\psi^2}{15}\right)}{1+\frac{4 \psi^2}{15}}-\nonumber\\
-\frac{8}{5} e^{-2 \psi} \psi-\frac{2
\psi^2}{1+\psi^2}.\end{eqnarray}
It is obvious that due to the large number of essential parameters
of the model, (4 parameters, $p_f, m,$ $\mu, q$), and initial
conditions, (2 independent  condi\-ti\-ons, $\Phi(0),$
$\dot{\Phi}(0)$), a consi\-de\-red cosmological model is extreme
rich in types of behavior. Therefore let us consider the main
ones. Here are the results of numerical solutions of Einstein and
Klein - Gordon equations for the degenerated Fermi system  in math
package Mathematica v7.
\begin{center}
\parbox{10cm}{\epsfig{file=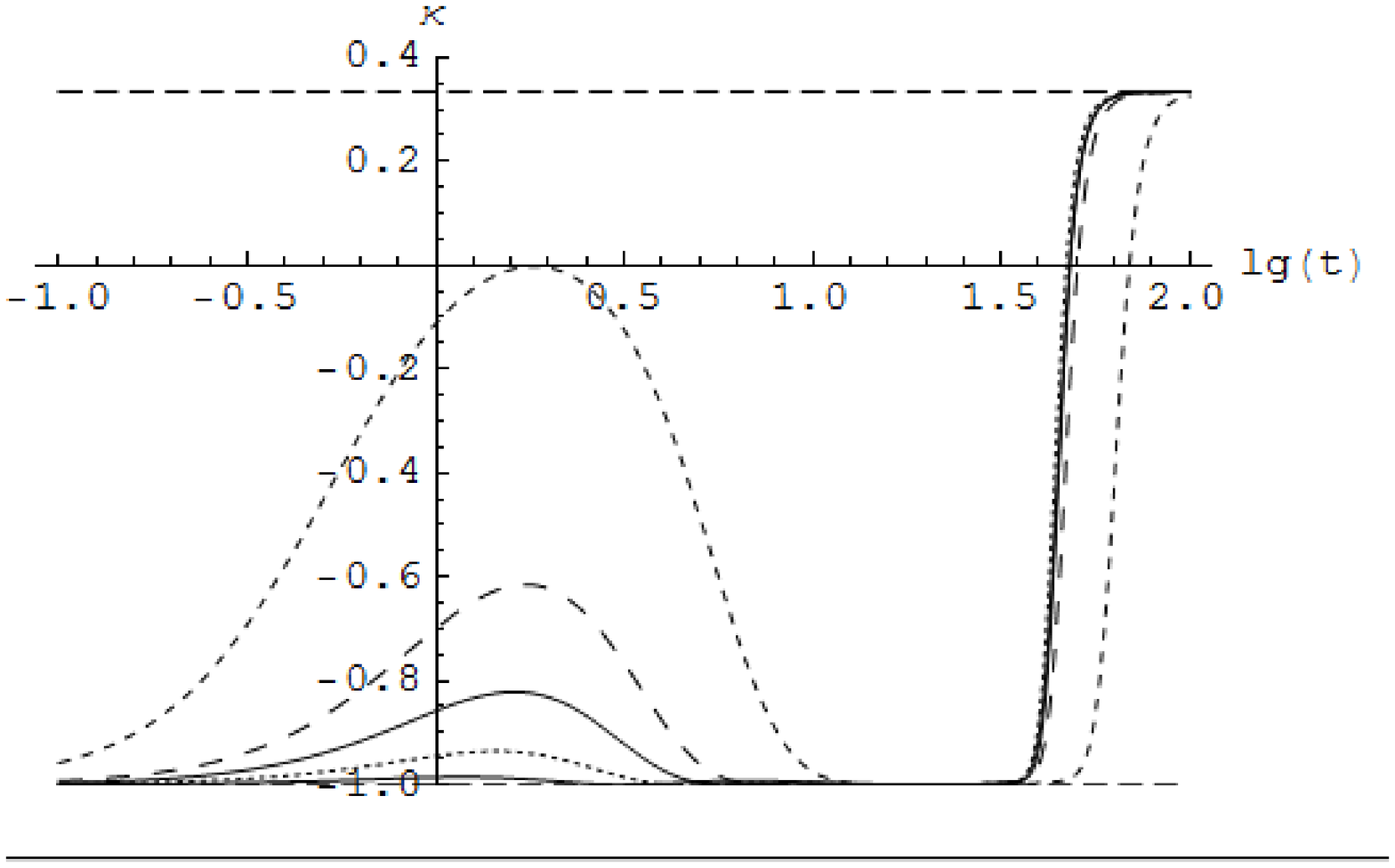, width=10cm,height=7cm}\vskip 11pt {\bf Fig.
1.} Evolution of the barotropic coefficient, $\varkappa$,
depending on mass of scalar bosons. Frequent dotted line:
$\mu=0,3$; dotted line: $\mu=0,1$; rare dotted line - $\mu=0,2$;
normal line: $\mu=0,25$; thick line: $\mu=0,35$. Everywhere:
$p_F=0,01; \hskip 6pt m=1; \hskip 6pt q=0,3; \Phi(0)=1; \hskip 6pt
\dot{\Phi}(0)=0.$ }
\end{center}
\begin{center}
\parbox{10cm}{\epsfig{file=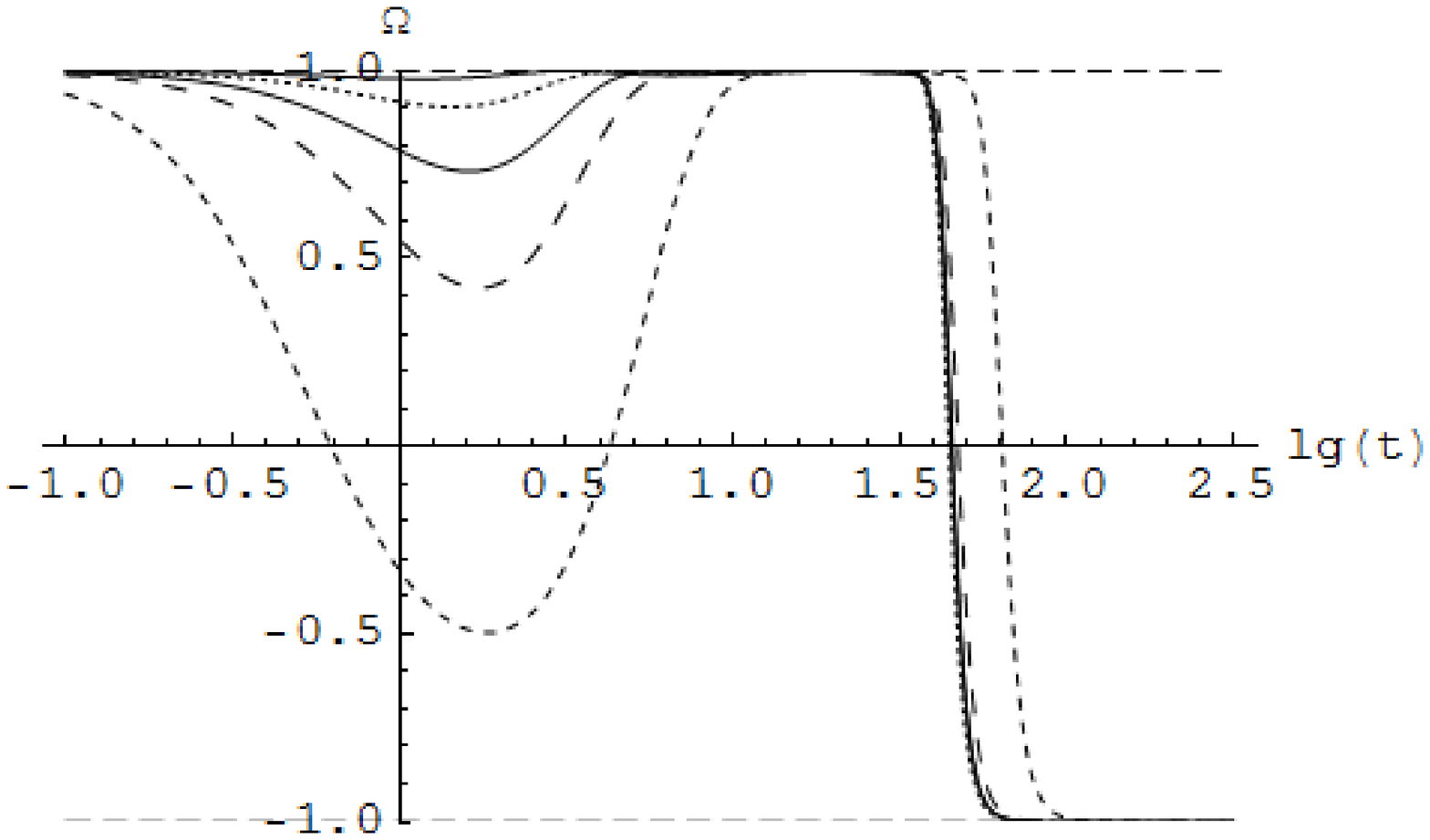, width=10cm,height=7cm}\vskip 11pt {\bf Fig.
2.} Evolution of the cosmological acceleration, $\Omega$,
depending on mass of scalar bosons. Frequent dotted line:
$\mu=0,3$; dotted line: $\mu=0,1$; rare dotted line: $\mu=0,2$;
normal line: $\mu=0,25$; thick line: $\mu=0,35$. Everywhere:
$p_F=0,01; \hskip 6pt m=1; \hskip 6pt q=0,3; \Phi(0)=1; \hskip 6pt
\dot{\Phi}(0)=0.$ }
\end{center}
\begin{center}
\parbox{10cm}{\epsfig{file=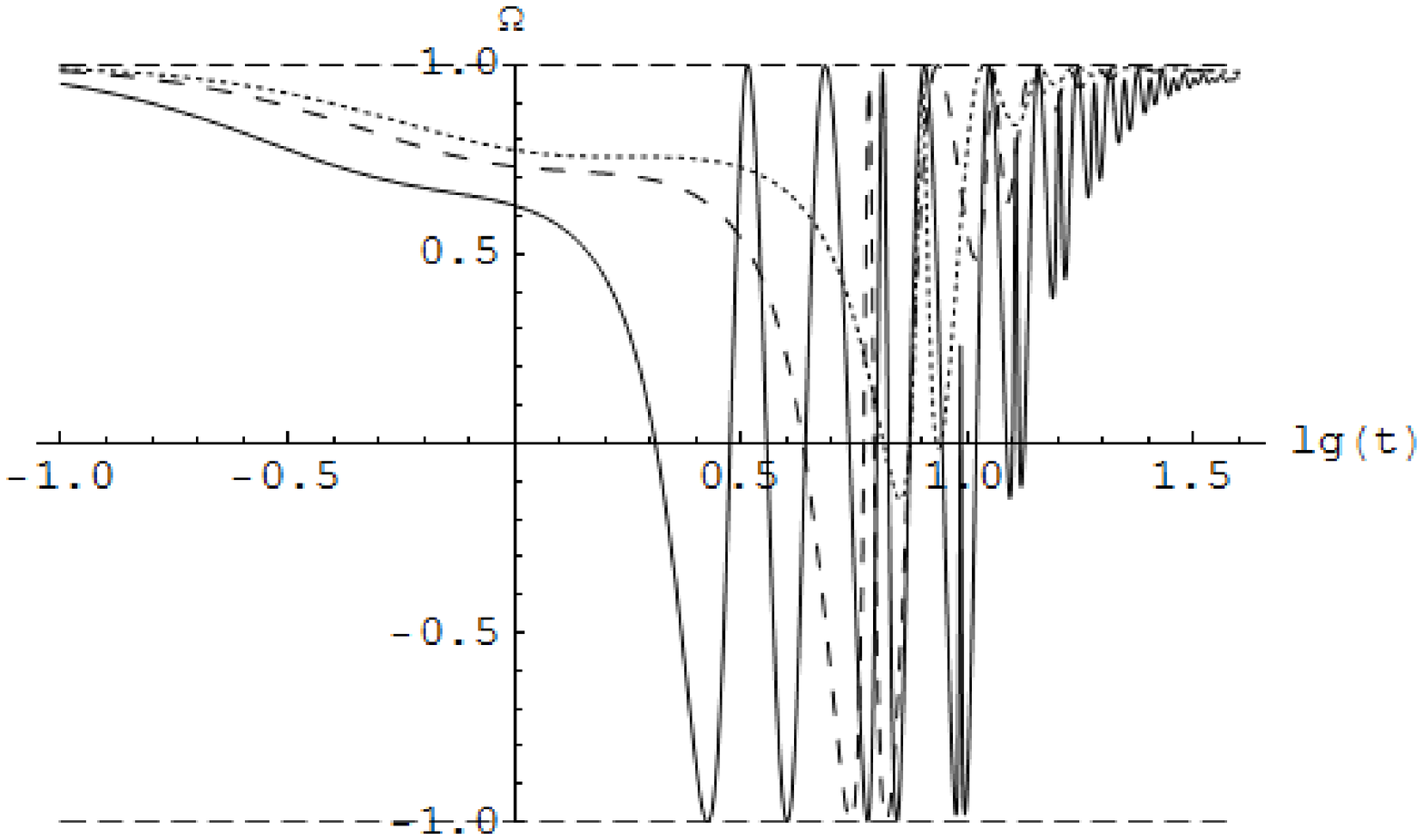, width=10cm,height=7cm}\vskip 11pt {\bf Fig.
3.} Evolution of the cosmological acceleration, $\Omega$,
depending on mass of scalar bosons. Frequent dotted line: $\mu=1$;
dotted line: $\mu=1,2$; normal line: $\mu=2$; thick line: $\mu=3$.
Everywhere: $p_F=0,01; \hskip 6pt m=1; \hskip 6pt q=0,3;
\Phi(0)=1; \hskip 6pt \dot{\Phi}(0)=0.$ }
\end{center}
Let us present the solutions which transgress to the accelerated
cosmological phase with maximum acceleration. These solutions
correspond to very small values of the fundamental charge.

\begin{center}
\parbox{10cm}{\epsfig{file=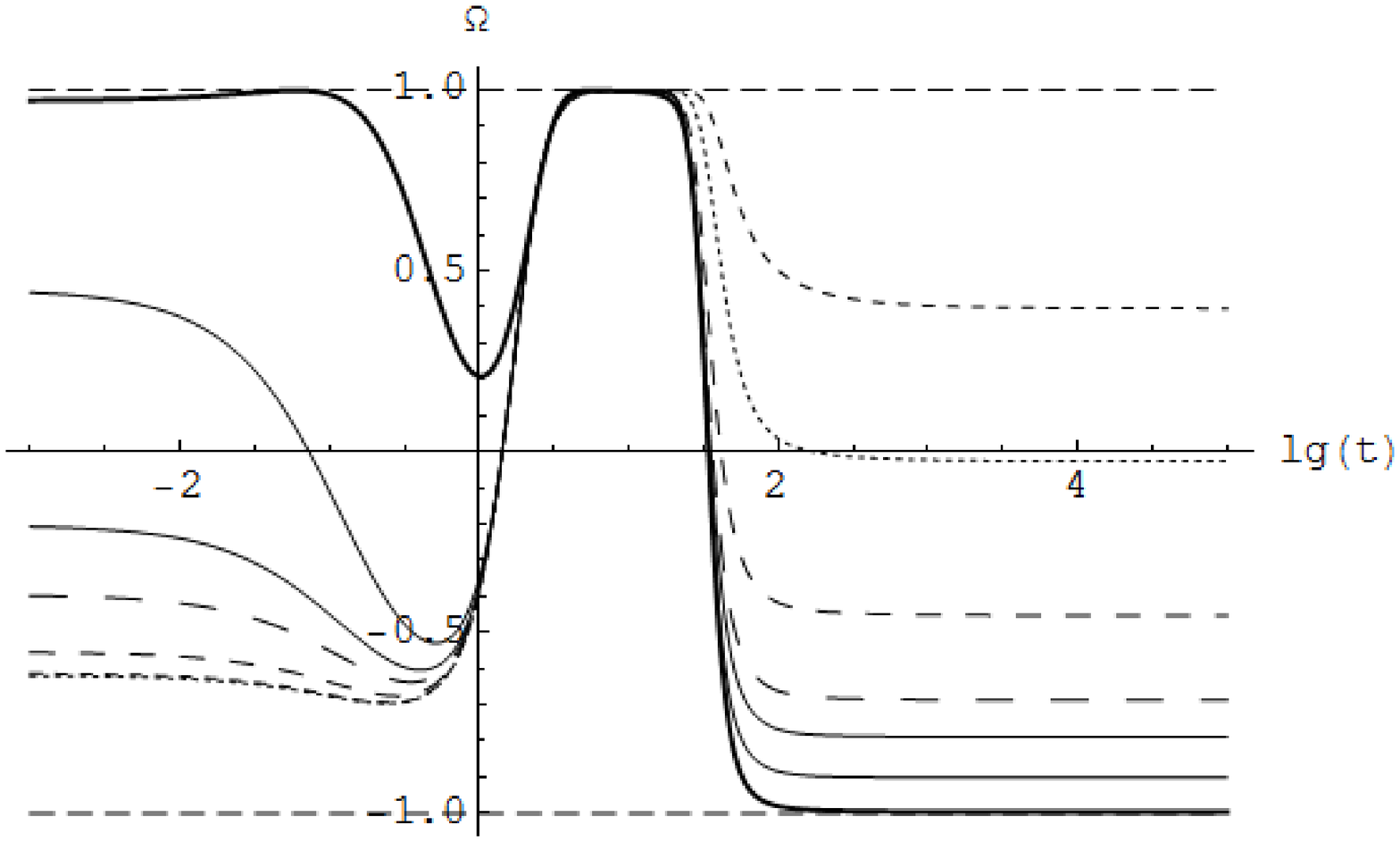, width=10cm,height=7cm}\vskip 11pt {\bf Fig.
4.} Evolution of the cosmological acceleration, $\Omega$,
depending on mass of fermions. Frequent dotted line: $m=0$; dotted
line: $m=0,001$; rare dotted line: $m=0,01$; very rare dotted
line: $m=0.03$; thin line: $m=0,05$; normal line: $m=0,1$, thick
line: $m=1$. Everywhere: $p_F=0,01; \hskip 6pt \mu=0,3; \hskip 6pt
q=1; \Phi(0)=1; \hskip 6pt \dot{\Phi}(0)=0.$ }
\end{center}
\begin{center}
\parbox{10cm}{\epsfig{file=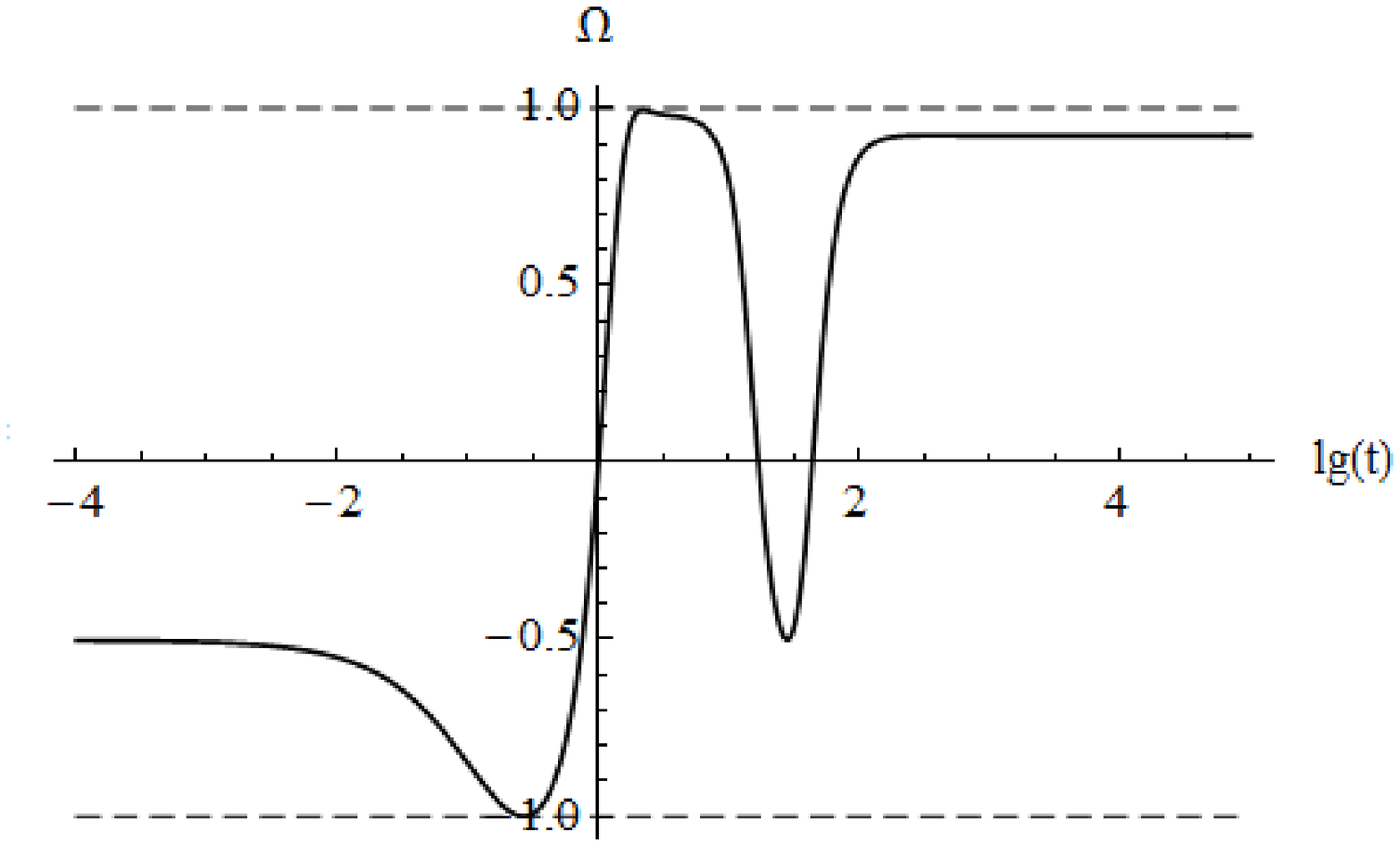, width=10cm,height=7cm}\vskip 11pt {\bf Fig.
5.} Evolution of the cosmological acceleration, $\Omega$ where
$p_F=0; \hskip 6 pt m=1; \hskip 6pt \mu=0,001; \hskip 6pt q\to
0$.}
\end{center}
\begin{center}
\parbox{10cm}{\epsfig{file=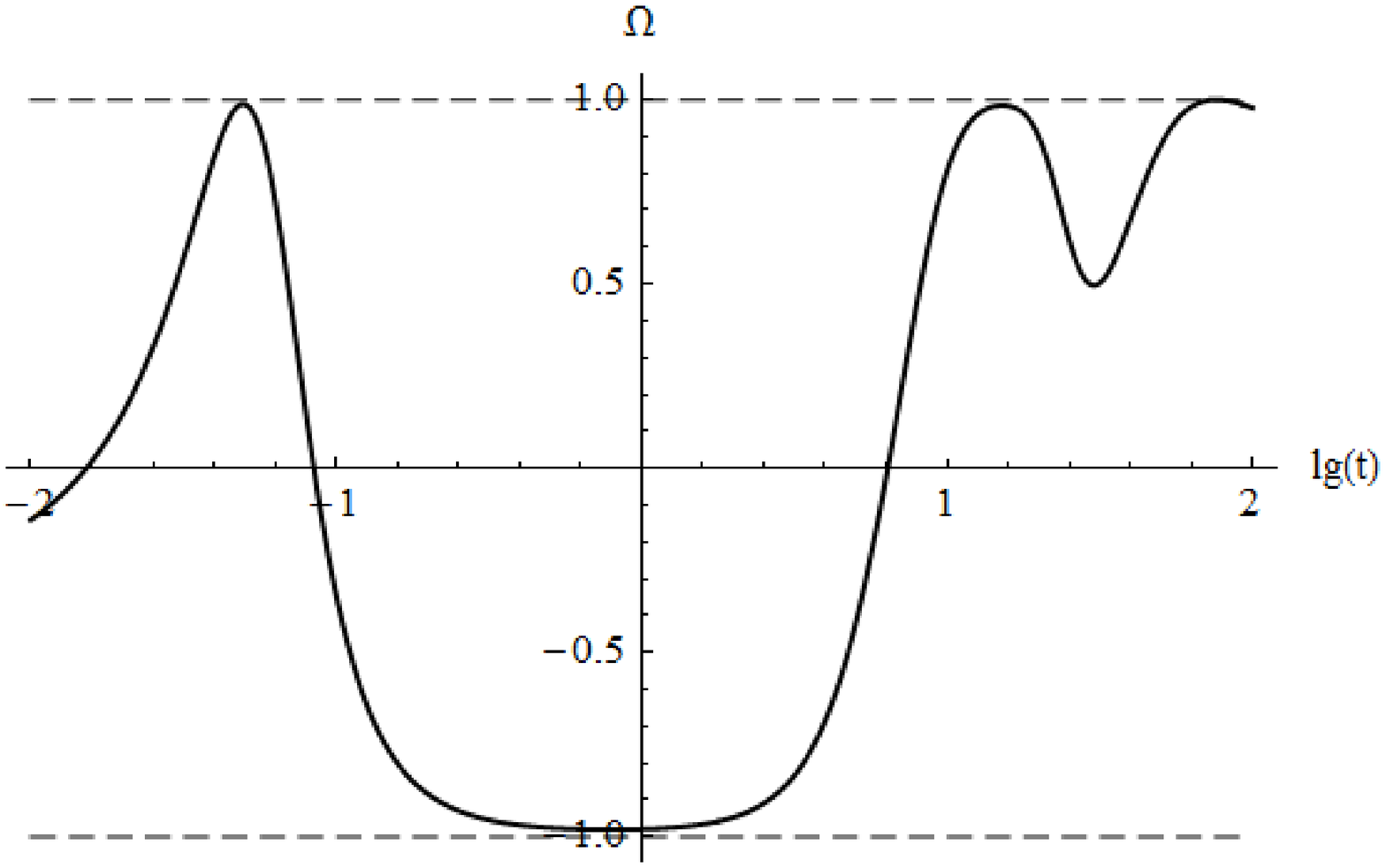, width=10cm,height=7cm}\vskip 11pt {\bf Fig.
6.} Evolution of the cosmological acceleration, $\Omega$ where
$p_F=0; \hskip 6 pt m=1; \hskip 6pt \mu=0,1; \hskip 6pt q\to 0$.}
\end{center}

\section{Conclusion}
Thus, we can resume:

\noindent%
1. Evolution of the cosmological model depends heavily on
the fundamental constants $q,\mu$ and $m$ has many types of behavior;\\
2. There are ranges of fundamental constants and initial
conditions, in which the cosmological models come out later on
stable acceleration;\\
3. Access to the phase later acceleration can be smooth or
accompanying by the acceleration os\-cil\-la\-ti\-ons; \\
4. Final acceleration is constant and can vary in the
range $-1 \leq \Omega \leq 1$.\\
Therefore, constant acceleration (deceleration)  is of a general
nature.\\
\\
Authors would like to thank Prof. Vitaly N. Melnikov, Prof. Dmitry
V. Gal’tsov and Prof. Andrey A. Grib for useful discussion.

\end{document}